\documentclass[aps,pre,twocolumn,superscriptaddress,groupedaddress]{revtex4}

\usepackage{graphicx}  
\usepackage{amsmath}
\usepackage{bm}        
\usepackage{amssymb}   

\usepackage{float}

\usepackage{color}
\newcommand{\BvdM}[1]{\textcolor{black}{#1}}

\begin{document}

 \title{Predicting the phase behaviour of mixtures of active spherical particles}
%
 \author{Berend van der Meer}
 \thanks{These two authors contributed equally}
 \author{Vasileios Prymidis}
 \thanks{These two authors contributed equally}
 \author{Marjolein Dijkstra}
  \author{Laura Filion}
  \email[Author to whom correspondence should be addressed: l.c.filion@uu.nl]{}
 
\affiliation{Soft Condensed Matter, Debye Institute for Nanomaterials Science,Utrecht University, Princetonplein 5, 3584 CC Utrecht, The Netherlands}

\begin{abstract}

An important question in the field of active matter is whether or not it is possible to predict the phase behaviour of these systems. Here, we study the phase coexistence of binary mixtures of torque-free active Brownian particles, for both systems with purely repulsive interactions and systems with attractions.  Using Brownian dynamics simulations, we show that phase coexistences can be predicted quantitatively for these systems by measuring the pressure and ``reservoir densities''. Specifically, in agreement with previous literature, we find that the coexisting phases are in mechanical equilibrium, i.e. the two phases have the same pressure. Importantly, we also demonstrate that the coexisting phases are in chemical equilibrium by bringing each phase into contact with particle reservoirs, and showing that for each species these reservoirs are characterized by the same density for both phases. Using this requirement of mechanical and chemical equilibrium we accurately construct the phase boundaries from properties which can be measured purely from the individual coexisting phases. This result highlights that  torque-free active Brownian systems follow simple coexistence rules, thus shedding new light on their thermodynamics.

\end{abstract}
 
\maketitle

\section{Introduction}
Experimental realizations of ``active'' colloidal particles, i.e. colloidal particles that self-propel, have opened the door to exploiting active building blocks in new colloidal systems (see e.g. \cite{marchetti2013hydrodynamics,aranson2013active,elgeti2015physics,bialke2015active,bechinger2016active,aubret2018targeted,bricard2013emergence,izri2014self}).   These active particles incessantly convert energy into self-propulsion and, as such, systems containing active particles are inherently out-of-equilibrium. 

Intriguingly, while active systems often exhibit behaviour fully prohibited in equilibrium systems, such as gas-liquid phase separation in purely repulsive systems   (see e.g. \cite{redner2013structure,cates2015motility}) and symmetry-breaking motion \cite{reichhardt2013active,di2010bacterial}, the steady-state behaviour of active systems can often be summarized by phase diagrams similar to their passive counterparts, i.e. consisting of single-phase regions and coexistence regions where the lever rule holds. For instance, fairly classic phase diagrams have been observed in the attraction-induced liquid-gas phase coexistence of active Lennard-Jones particles~\cite{prymidis2015self,prymidis2016vapour}, the melting and motility-induced phase separation observed in repulsive, self-propelled spheres~\cite{henkes2011active,redner2013structure,digregorio2018full,klamser2018thermodynamic,paliwal2019role,siebert2017phase}, dumbbells~\cite{cugliandolo2017phase}, and even polygons~\cite{prymidis2016state,ilse2016surface}, as well as binary mixtures \cite{trefz2016activity,kolb2020active,van2016removing,dolai2018phase}. 
In equilibrium, phase boundaries and coexistences are inherently tied to bulk thermodynamic properties. Since coexisting phases have equal pressures and equal chemical potentials, the bulk properties of the individual phases provide a direct route to predicting phase coexistences - a strategy commonly used to draw phase diagrams.  However, for active particles we are still in the process of developing thermodynamic frameworks that describe their phase boundaries (see e.g \cite{takatori2015towards,solon2015pressure,takatori2014swim,solon2015pressure2,bialke2015negative,
falasco2015mesoscopic,menzel2016dynamical,speck2016ideal,winkler2015virial,solon2018generalized,paliwal2018chemical,krinninger2016nonequilibrium,hermann2019phase,omar2019microscopic}).
 
Early attempts  to perform such a task include the work of Takatori and Brady~\cite{takatori2015towards} and the work of Solon \textit{et al.}~\cite{solon2015pressure}. These attempts were motivated by the existence of  a mechanical pressure for spherical active particles  \cite{takatori2014swim,solon2015pressure2,bialke2015negative,falasco2015mesoscopic,speck2016ideal,winkler2015virial} but led to only qualitative predictions of the phase diagrams of the systems under study. In the former work, phase coexistences were predicted using  an approximate  generalized free energy for the system, while in the latter work it was predicted by means of a Maxwell construction, which was ultimately found to not apply to active matter systems.  More recently, Solon \textit{et al.}~\cite{solon2018generalized} developed a  theory of phase-separating active particles starting from a generalized Cahn-Hilliard description.  Interestingly, in this description, their prediction of the binodals requires information not only on the bulk coexisting phases, but also on the interface between them.  Other attempts to explain the thermodynamics of active particles include methods associated with power functional concepts \cite{krinninger2016nonequilibrium,hermann2019phase}. In this direction, Hermann  {\it et al.} \cite{hermann2019phase} developed a microscopic theory for bulk and interfacial behavior in these systems and were able to capture the phase diagram of an active Brownian particle system using only a few fit parameters. However, despite these, and several other attempts to use thermodynamical concepts to predict active phase separation, a complete understanding of the applicability of such concepts is still lacking.

In this paper, we attack the problem from a different angle, and explore the possibility of chemical and mechanical equilibrium in these systems as directly as possible using computer simulations.  Specifically, we take the existence of a mechanical pressure for torque-free spherical active particles as a starting point~\cite{solon2015pressure,solon2015pressure2} and introduce a numerical method to probe the possibility of chemical equilibrium in the system. In contrast to previous literature \cite{tailleur2008statistical,stenhammar2013continuum,wittkowski2014scalar,marconi2015towards,takatori2015theory,paliwal2018chemical,guioth2018large,hermann2019phase}, our aim is not to measure a chemical potential, but simply to test whether there is a quantity, similar to a chemical potential, that is equal in both phases -- which we can measure in simulations where there is no direct contact between the coexisting phases. To this end, we develop a simulation method which lets us measure a quantity which is analogous to the chemical potential in equilibrium systems, namely a reservoir density. In classical statistical physics the chemical potential arises when we connect a system to a particle reservoir. The chemical potential is then directly related to the density of the reservoir.  Here, we copy this idea, and develop a simulation method that measures these reservoir densities instead of the chemical potential. We then explore the possibility of chemical and mechanical equilibrium in active systems by examining two different systems: an out-of-equilibrium mixture of passive and active attractive particles, and an active-active mixture of purely repulsive particles. As expected (e.g \cite{solon2015pressure2}) we find mechanical equilibrium for the coexisting phases. More importantly, we show that the reservoirs of both species are characterized by the same density for coexisting phases. Finally, we show explicitly that the phase coexistence can be \textit{predicted} quantitatively by measuring the pressure and the reservoir densities.

\section{Model and Methods}
\subsection{Model systems}
In this paper we investigate phase coexistence in two different types of active systems. The first system consists of a mixture of active and passive attractive particles. Here, phase separation is induced by the attractions. Thus at mild activity, we are simply perturbing the attraction-dominated liquid-gas phase separation. In the second system, we study a purely out-of-equilibrium phenomenon, namely motility-induced phase separation in a mixture of repulsive particle with different amounts of self-propulsion. \BvdM{By choosing these two systems we explore both the case where the activity only perturbs the phase behaviour and where the activity is largely responsible for the phase coexistence, ensuring that our results are as general as possible for torque-free systems. Moreover, these systems are excellent starting points as phase coexistences have been studied quite extensively in their single-component counterparts (see e.g. ~\cite{prymidis2015self,prymidis2016vapour,cates2015motility}).}

\subsubsection{Lennard-Jones active-passive mixture}
For the mixture of active and passive attractive particles we consider a three-dimensional system of $N$ spherical particles that interact via the well-known Lennard-Jones potential:
\begin{equation}
  \beta U(r)= 
  4\beta \epsilon_{LJ} \left( \left(\frac{\sigma}{r}\right)^{12}-\left(\frac{\sigma}{r}\right)^{6} \right)
\end{equation}
truncated and shifted at $r_c=2.5\sigma$ with $r=|\boldsymbol{r}_i-\boldsymbol{r}_j|$ the center-of-mass distance between particle $i$ and $j$, $\sigma$ the particle diameter, $\beta \epsilon_{LJ}$ the energy scale, and $\beta=1/k_BT$, where $k_B$ is the Boltzmann constant and $T$ is the temperature. Out of the $N$ particles, we ``activate'' a subset of $N_a$ particles by introducing a constant self-propulsion force $f_a$ along the self-propulsion axis $\hat{\textbf{u}}_i$.  We denote the fraction of active particles by $x = N_{a} /N$.  The total density of the system is given by $\rho = N/V$, where $V$ is the volume of the system.

\subsubsection{Weeks-Chandler-Andersen active-active mixture}

For the active-active mixture of repulsive particles we consider a two-dimensional system of $N$ particles interacting via the Weeks-Chandler-Andersen potential:
\begin{equation}
  \beta U(r)= 
  4\beta \epsilon_{WCA} \left( \left(\frac{\sigma}{r}\right)^{12}-\left(\frac{\sigma}{r}\right)^{6} + \frac{1}{4}\right),
\end{equation}
with the interaction cutoff radius $r_c=2^{1/6}\sigma$ and the energy scale $\beta\epsilon_{WCA}$. Specifically, we consider an active-active mixture with the self-propulsions of fast and slow species being $f_f$ and  $f_s$, respectively.  We denote the fraction of fast particles by $x = N_{f} /N$ with $N_f$ the number of fast particles. The total density of the system is given by $\rho = N/A$, where $A$ is the area of the system. Note that in this case we find a phase separation between a high density ``crystal''  phase and a low density gas phase. 

\subsection{Dynamics}
These systems are simulated using overdamped Brownian dynamics. Specifically, the equations of motion for particle $i$ are:
\begin{eqnarray}
\dot{{\bf r}}_i(t) &=& \beta D_0 \left[  - \nabla_i U (t)+  f_i\hat{{\bf u}}_i(t)\right] + \sqrt{2D_0} {\bm \xi}_i(t)\label{eqtr}\\
{\dot{\hat {\bf u}}}_i (t) &=& \sqrt{2 D_r} \hat{\bf u}_i (t)\times  {\bm \eta}_i(t)\label{eqor},
\end{eqnarray} 
where $\bm{\xi}_i(t)$  and $\bm{\eta}_i(t)$ are stochastic noise terms with zero mean and unit variance, and $f_i$ is the self-propulsion force.  Note that for passive particles $f_i  = 0$. The translational diffusion coefficient $D_0$ and the rotational diffusion constant $D_r$ are linked via the Stokes-Einstein relation $D_r=3D_{0}/\sigma^2$.  We measure time in units of the short-time diffusion $\tau=\sigma^2/D_0$. 

\subsection{The normal component of the local pressure tensor}

For systems of self-propelled particles that do not experience torque, such as those we consider here, it has been shown that  the mechanical  pressure  is a state function~\cite{solon2015pressure2}. Moreover,  when such a system undergoes phase separation, the coexisting phases have the same bulk pressure~\cite{bialke2015negative}. Here, we use an expression for the local pressure of an active system with only isotropic interactions~\cite{paliwal2017non}, which we generalize to binary mixtures.
Our expression reproduces the pressure for systems of a single species~\cite{solon2015pressure}, and, when spatially  averaged  over the whole system, also amounts to the known expressions for the pressure of non-confined systems~\cite{winkler2015virial, falasco2015mesoscopic}.

Consider a binary system in $d$ dimensions with $N_1$ particles of species $1$ and $N_2$ particles of species $2$, whose equations of motion are given by Equations \ref{eqtr} and \ref{eqor}. These equations should now be viewed with the appropriate generalization in $d$ dimensions and with $U(r)$ being an arbitrary pair-wise potential, with $r=|\boldsymbol{r}_i-\boldsymbol{r}_j|$ the center-of-mass distance between particle $i$ and $j$. The only difference between the two species is the value of the self-propulsion force, such that $f_i=f^{(1)}$ for $i\in N_1$  and $f_j=f^{(2)}$ for $j\in N_2$, with $f^{(1)}\neq f^{(2)}$. We also define the angle between the axis of self-propulsion $\hat {\bf u} (t)$ and a fixed coordinate system as $\Omega(t)$. 

Let us now define the microscopic density field for the two species, which we denote as $\Psi^{(1)}(\boldsymbol{r},\Omega)$ and $\Psi^{(2)}(\boldsymbol{r},\Omega)$, such that
\begin{align}
 \Psi^{(1)}(\boldsymbol{r},\Omega)&=\sum_{i=1}^{N_1}\delta(\boldsymbol{r}-\boldsymbol{r}_i)\delta(\Omega-\Omega_i),\\
 \Psi^{(2)}(\boldsymbol{r},\Omega)&=\sum_{i=1}^{N_2}\delta(\boldsymbol{r}-\boldsymbol{r}_i)\delta(\Omega-\Omega_i).
\end{align}

For later convenience, we define the moments 
\begin{align}
 \rho^{(\alpha)}(\boldsymbol{r})&=\int d\Omega  \Psi^{(\alpha)}(\boldsymbol{r},\Omega), \label{eq:density}\\
 \boldsymbol{m}^{(\alpha)}(\boldsymbol{r})&=\int d\Omega \hat{{\bf u}}(\Omega)  \Psi^{(\alpha)}(\boldsymbol{r},\Omega), \label{eq:polarization}\\
  \boldsymbol{s}^{(\alpha)}_{ij}(\boldsymbol{r})&=\int d\Omega  \hat{{\bf u}}_i(\Omega) \hat{{\bf u}}_j(\Omega)   \Psi^{(\alpha)}(\boldsymbol{r},\Omega),
\end{align}
where $i,j$ denote the spatial components of the corresponding vectors and tensors. The field $\rho^{(\alpha)}(\boldsymbol{r})$ is  the local density of species ${\alpha}$ while the field $ \boldsymbol{m}^{(\alpha)}(\boldsymbol{r})$ is the corresponding local polarization.

We assume that the system is confined in the $\hat{z}$ direction by a wall  while it is translationally invariant in the dimensions perpendicular to $\hat{z}$, \textit{i.e.} periodic boundary conditions are applied. Following Refs.~\cite{solon2015pressure, solon2015pressure2},	 the pressure felt by the wall can be written as 
\begin{align}\label{eq:wallpressure}
 P_{\rm wall}=P_{\rm  id}(z)+P_{\rm  vir,\hat{z}}(z)+\sum_\alpha P^{(\alpha)}_{\rm  swim,\hat{z}}(z),
\end{align}
where $z$ denotes any point in the bulk of the system,
\begin{align}\label{eq:ideal}
 P_{\rm id}(z)=\left\langle\rho(z)\right\rangle k_BT=\frac{1}{L^{d-1}}\int d\boldsymbol{r}^{d-1} \left\langle\rho(\boldsymbol{r})\right\rangle k_BT,
\end{align}
is the ideal component of the pressure, where we have spatially integrated over the dimensions that are perpendicular to the $\hat{z}$ dimension and divided by the surface $L^{d-1}$ that we integrated over, $\rho(\boldsymbol{r})=\sum_\alpha  \rho^{(\alpha)}(\boldsymbol{r})$ is the total density at point $\boldsymbol{r}$, and brackets denote an average at the steady state over noise realizations,
\begin{eqnarray}
&&  P _{\rm vir, \hat{z}}(z)=  \\ &&
  \frac{1}{L^{d-1}}\int_{z^{\prime\prime}<z}d\boldsymbol{r}^{\prime\prime}
  \int_{z^{\prime}>z}d\boldsymbol{r}^{\prime} \left\langle\rho(\boldsymbol{r}^{\prime\prime})\rho(\boldsymbol{r}^{\prime})\right\rangle\partial_{\hat{z}} U(\left|\boldsymbol{r}^{\prime}-\boldsymbol{r}^{\prime\prime}\right|) ,\nonumber
\end{eqnarray}
is the standard local virial term, and
\begin{eqnarray}\label{eq:swim}
 P^{(\alpha)}_{\rm swim, \hat{z}}(z)= \frac{ D_0  f^{(\alpha)}}{(d-1)D_r} 
 \left[-\beta\partial_{\hat{z}}  U(z)  \left\langle\boldsymbol{m}^{(\alpha)}_{\hat{z}}(z)\right\rangle  + \right. \nonumber \\ \left.
 \beta f^{(\alpha)} \left\langle\boldsymbol{s}^{(\alpha)}_{{\hat{z}}{\hat{z}}}(z)\right\rangle- \partial_{\hat{z}}\left\langle\boldsymbol{m}^{(\alpha)}_{\hat{z}}(z)\right\rangle\right]
\end{eqnarray}
is the local swim pressure of species $\alpha$. In Equation \ref{eq:swim} $U(z), \boldsymbol{m}^{(\alpha)}_{\hat{z}}(z)$ and $ \boldsymbol{s}^{(\alpha)}_{{\hat{z}}{\hat{z}}}(z)$ have been averaged similarly to Equation \ref{eq:ideal}. 

Now, since the right-hand-side of Equation \ref{eq:wallpressure} is a local quantity that does not depend on any wall properties, we straightforwardly define the normal component of the local pressure of the system as 
\begin{align}\label{eq:normal}
 P_{\rm N}(z)=P_{\rm  id}(z)+P_{\rm  vir,\hat{z}}(z)+\sum_\alpha P^{(\alpha)}_{\rm  swim,\hat{z}}(z).
\end{align}
In our particle simulations, we divide the simulation box into slabs, and measure the normal component of the pressure for each slab. The contributions to each slab of the ideal and swim components of the pressure  can be calculated straightforwardly, while for the virial component we follow Ref.~\cite{ikeshoji2003molecular}.

\subsection{Reservoir simulations}~\label{sec:ressim}
 The standard method for showing chemical equilibrium in passive systems is to measure the chemical potential in both phases, for both species. For a passive system, this can be done in a number of different ways depending on the exact circumstance - ranging from e.g. grand canonical simulations to thermodynamic integration~\cite{frenkel2001understanding}. However, for systems containing active particles these methods do not directly apply, and hence we will address the question following a different route. 
 
In a textbook derivation of the chemical potential, one typically attaches the system in question to a large particle reservoir, and allows the particles of a given species to travel between the subsystem in question, and the particle reservoir. The subsystem is then in a grand-canonical ($\mu VT$) ensemble, with $\mu$ set by the chemical potential of the reservoir. Hence, if two systems have the same chemical potential, one must be able to connect them to the same particle reservoir, i.e. one with the same particle density. Here, we follow a similar procedure with our simulations.

\begin{figure*} 
\begin{center}
\includegraphics[width=0.7\textwidth]{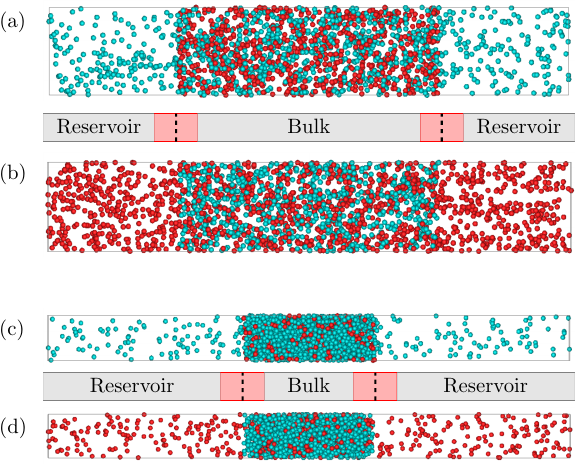} 
\end{center}
\caption{
The coexisting  gas phase of a Lennard-Jones active-passive mixture in contact with (a) a passive particle reservoir and with (b) an active particle reservoir. The coexisting liquid phase of a Lennard-Jones active-passive mixture in contact with (c) a passive particle reservoir and with (d) an active particle reservoir. Note that the reservoirs associated with the gas (a,b) and liquid (c,d) phase  have different cross-sections, making the reservoir densities in (a,b) seem a lot higher compared to those in (c,d) -- even though they are equal. 
}
\label{resdensOVERVIEW}
\end{figure*}

Specifically, we connect a binary phase  to particle reservoirs that only contain a single species. 
To this end, we divide our simulation box into two sections, one which contains the ``bulk'' binary phase, with the other part of the box acting as a passive (or active) particle reservoir (see Figure \ref{resdensOVERVIEW}).  We place a semi-permeable membrane at the division which only allows one of the two species, which we call species $R$, to pass through at no energy cost. For the other species, the wall is impenetrable with the wall-particle interaction given by the purely repulsive Weeks-Chandler-Andersen-like wall potential:
\begin{equation}
  \beta U(z)= 
  4\beta \epsilon_{WCA} \left( \left(\frac{\sigma}{z}\right)^{12}-\left(\frac{\sigma}{z}\right)^{6} + \frac{1}{4}\right),
\end{equation}
where $z$ is the distance of a particle in the bulk to the nearest semi-permeable wall, $\beta \epsilon_{WCA} = 40$, and the interaction is cut off at a distance $z = 2^{1/6}\sigma$.

At the start of the reservoir simulation, the bulk binary phase in the middle section of the box has the desired partial densities of the two different species, and the reservoir is initialized with a gas of species $R$ of arbitrary density. Over the course of the simulation, particles of species $R$ will exchange between the bulk binary phase and the reservoir. Hence, the density of this species will change in both the bulk phase, as well as in reservoir. Additionally, a small fraction of the bulk phase typically builds up on the semi-permeable membrane, and sometimes depletes the bulk region of the confined species.

To counteract such deviations from the desired partial densities in the bulk phase we ``tune'' the density of both species during the equilibration of the simulation. This tuning is realized by either adding or removing particles. Specifically, we measure at fixed intervals the partial densities of each species in the center of the bulk phases (i.e. away from the semi-permeable membrane).  If the partial density of one of the species is too high, particles of this species are removed randomly. If the partial density of one of the species is too low, particles of this species are added randomly. Eventually, after repeating this procedure many times, the average partial densities of both species in the bulk region reach their desired constant, and a steady state reservoir density is obtained. At this point, we stop removing and adding particles, and check that the reservoir density remains constant. In this way we have checked that the binary phase and the particle reservoir are in chemical equilibrium.

\section{Results}
\subsection{Lennard-Jones active-passive mixture}

To start our investigation we construct a liquid-gas coexistence in an active-passive mixture, in the regime where the system phase separates due to attractions~\cite{prymidis2015self}. Here, we set the energy scale $\beta\epsilon_{LJ}=1.2$, which for a purely passive system ($x = 0$) results in a well-characterized liquid-gas coexistence at intermediate densities~\cite{smit1992phase}. In the following we fix the self-propulsion force $f_a = 10 k_BT /\sigma$ for all active particles, and perform simulations for a range of values of the active fraction $x$ and overall system density $\rho$ in an elongated simulation box. 

 \begin{figure*}
\begin{center}
\includegraphics[width=0.7\textwidth]{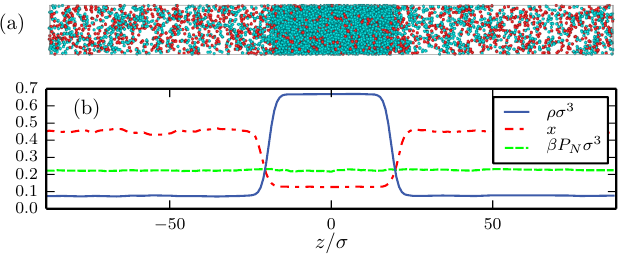} 
\end{center}
\caption{(a) Liquid-gas coexistence of a Lennard-Jones active-passive mixture with an overall active fraction $x=0.22$ at an overall density $\rho \sigma^3=0.20$. For the active particles the self-propulsion force equals  $f_a=10 k_B T / \sigma$. Note that active (passive) particles are coloured red (blue). (b) The corresponding density, composition, and pressure profile perpendicular to the planar interfaces. The system contains $N=8192$ particles.}
\label{profile}
\end{figure*}

We study the phase diagram for this mixture via direct coexistence simulations. We simulated approximately 8000 particles in a three-dimensional elongated box with dimensions $L_{\hat{z}}=12L_{\hat{y}}=12L_{\hat{x}}$. \BvdM{Note that the use of an elongated box both ensures that the interface is on average planar, and minimizes the number of particles near the interface.} For the highest density (smallest simulation box) we consider, the dimensions of the box are approximately $12 \sigma \times 12 \sigma \times 144\sigma$. Thus, the short axis of the box is much larger than the persistence length of the active particles, which is $ \beta D_0 f/2D_r \approx 1.67  \sigma$. Note that we performed direct coexistence simulations for system compositions $x\in[0,0.4]$, and total densities $\rho\sigma^3=0.20, 0.30$ and $0.40$. These simulations were initiated from a configuration where all particles are located within a dense slab  and ran for approximately  $3500\tau$. We collected data  only for the last $500\tau$.

\begin{figure*}[htb!]
\begin{center}
\includegraphics[width=1\textwidth]{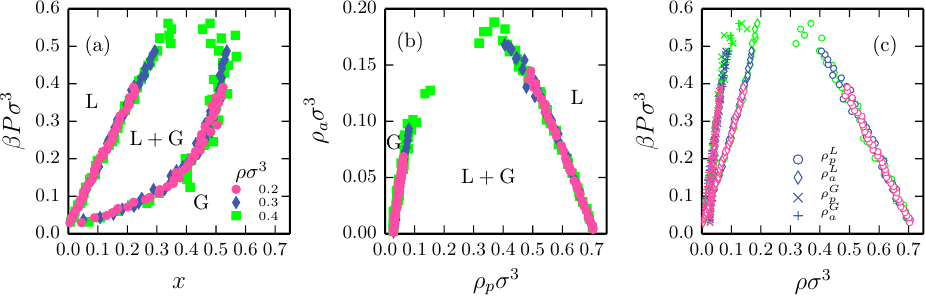} 
\end{center}
\caption{(a) Phase diagram of the Lennard-Jones active-passive mixture in the reduced pressure-composition $P$-$x$ representation. (b) The same phase diagram in the reduced active particle number density-passive particle number density $\rho_a$-$\rho_p$ representation. (c) Coexistence lines in the $P$-$\rho_a$ and $P$-$\rho_p$ representations.}
\label{phasediagLJ}
\end{figure*} 

In Figure \ref{profile}(a), we show a typical snapshot of an active-passive mixture exhibiting a liquid-gas coexistence. Here, the active and passive particles are plotted as red and blue, respectively. In Figure \ref{profile}(b), we plot the corresponding density $\rho(z)$, active particle fraction $x(z)$, and the normal pressure  $P_N(z)$ along the long axis of the box.  Note that in the bulk regime of either phase, this normal pressure $P_N$ will be equal to the bulk pressure $P$ of the phase in question.  Figure \ref{profile}(b) shows that, in this case, the system exhibits a gas-liquid coexistence with the gas characterized by density $\rho^G$ and composition $x^G$, and the liquid characterized by density $\rho^L$  and composition $x^L$. We always find the gas phase to be more rich in active particles, which is reminiscent of segregation phenomena seen in other studies of active mixtures~\cite{mccandlish2012spontaneous,stenhammar2015activity}.
Note that in all of our simulations, the pressure is the same in both coexisting phases indicating that the system is in mechanical equilibrium.  

Using composition and pressure profiles, similar to those shown in Figure \ref{profile}(b), we map out the coexisting compositions and pressures of our active-passive mixtures, for a wide range of overall system densities  $\rho$ and compositions $x$. The results are plotted in Figure \ref{phasediagLJ}(a). Similar to passive systems, we find that the phase behaviour collapses in this representation, i.e. the lever rule holds within the coexistence region.  
%
%

The validity of the lever rule is also evident when plotting the phase diagram in the active density - passive density ($\rho_a$-$\rho_p$) representation, as shown Figure \ref{phasediagLJ}(b). This is consistent with the simulation results of Ref.~\cite{trefz2016activity} where they summarized the phase behaviour of a different active-passive mixture in the $\rho_a$-$\rho_p$ representation. We also investigate the pressure dependence of the partial densities of each species in the coexisting liquid and gas phases.  In Figure \ref{phasediagLJ}(c) we plot these partial densities $\rho_i^\gamma$ vs. pressure $P$ with $i$ denoting the species (active or passive) and $\gamma$ denoting the phase (liquid(L) or gas(G)). 

\begin{figure} 
\begin{center}

\includegraphics[width=0.5\textwidth]{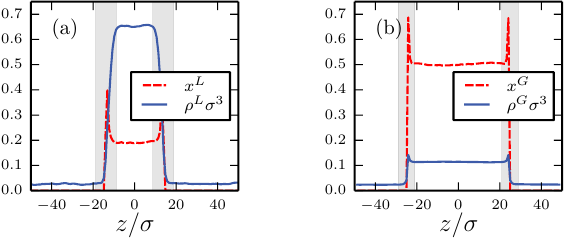}
\end{center}
\caption{(a,b) Density and composition profiles of the coexisting liquid (a) and gas (b) phase of a Lennard-Jones active-passive mixture in contact with a passive particle reservoir. The partial densities of the binary phases were chosen to correspond to the coexistence at pressure  $\beta P \sigma^3=0.34$. Note that some active particles adsorb at the wall. We thus exclude these interfacial regions (shaded areas) in the determination of the bulk density and composition, as well as in the determination of the reservoir density.
}
\label{resdensINT}
\end{figure}

Our results so far clearly demonstrate that the pressure is a key variable in controlling phase coexistences in our active-passive mixture: all phase coexistences are characterized by equal bulk pressures in the two phases. However, phase coexistence in an equilibrium binary system requires not only equal pressures between the two phases, but also equal chemical potentials for both species. This raises the question whether we can identify a quantity analogous to the chemical potential in active-passive mixtures which similarly controls the phase coexistence.


To this end, we perform reservoir simulations on these liquid and gas phases, with the other part of the box acting as a passive (or active) particle reservoir (see Figure \ref{resdensOVERVIEW}), as described in Section~\ref{sec:ressim}. Specifically, we select a binary gas and liquid that coexist, and connect them to particle reservoirs that only contain a single species.   Note that in total we will need four simulations per coexistence point, namely: the gas in contact with a passive particle reservoir, the gas in contact with an active particle reservoir, the liquid in contact with a passive particle reservoir, and the liquid in contact with an active particle reservoir. This is shown in Figure \ref{resdensOVERVIEW}.  
The goal will be to determine whether the active (and passive) reservoirs associated with the  coexisting phases are the same. If they are, then we can infer that they are in chemical equilibrium with each other.

During the equilibration of these simulations, typically a small fraction of the bulk particles builds up on the semi-permeable membrane, and sometimes depletes the bulk region of the confined species.  To counteract this adsorption (see Figure \ref{resdensINT}) and ensure that the bulk phase has the correct density and composition far away from the semi-permeable membrane, we tune the number of particles of both species  during equilibration. Eventually, the average partial densities of both species in the bulk region reach their desired value, and also the density of the reservoir reaches a constant value.

\begin{figure}[t!] 
\includegraphics[width=0.5\textwidth]{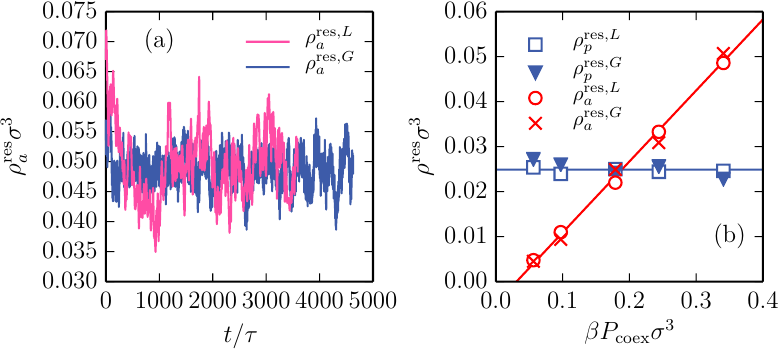}
\caption{(a) The density in the active particle reservoir over time for both the liquid and the gas. Both phases converge to the same reservoir density of active particles. (b) The reservoir densities as a function of the coexistence pressure $P_{\mathrm{coex}}$. The liquid and gas phase are in contact with the same reservoirs, showing that there is chemical equilibrium between the phases, for each species.}
\label{resdenstime}
\end{figure}

In Figure \ref{resdenstime}(a), we show the time evolution of the densities in the active particle reservoirs for the coexisting liquid and gas phases at $\beta P\sigma^3=0.34$. Note that although we chose a high initial density of the reservoirs in both cases, both reservoir densities quickly converged to the same density. In Figure \ref{resdenstime}(b), we plot the densities of both the active and passive reservoirs as a function of the coexistence pressure $P_{\mathrm{coex}}$. Clearly, for all coexisting liquid-gas pairs we find the same reservoir densities: $\rho_p^{\mathrm{res},L}=\rho_p^{\mathrm{res},G}$ and $\rho_a^{\mathrm{res},L}=\rho_a^{\mathrm{res},G}$. Hence, the two coexisting phases can be thought of being in chemical equilibrium with the same reservoir.

\subsection{Weeks-Chandler-Andersen active-active mixture}

\begin{figure*} 
\begin{center}
\includegraphics[width=0.7\textwidth]{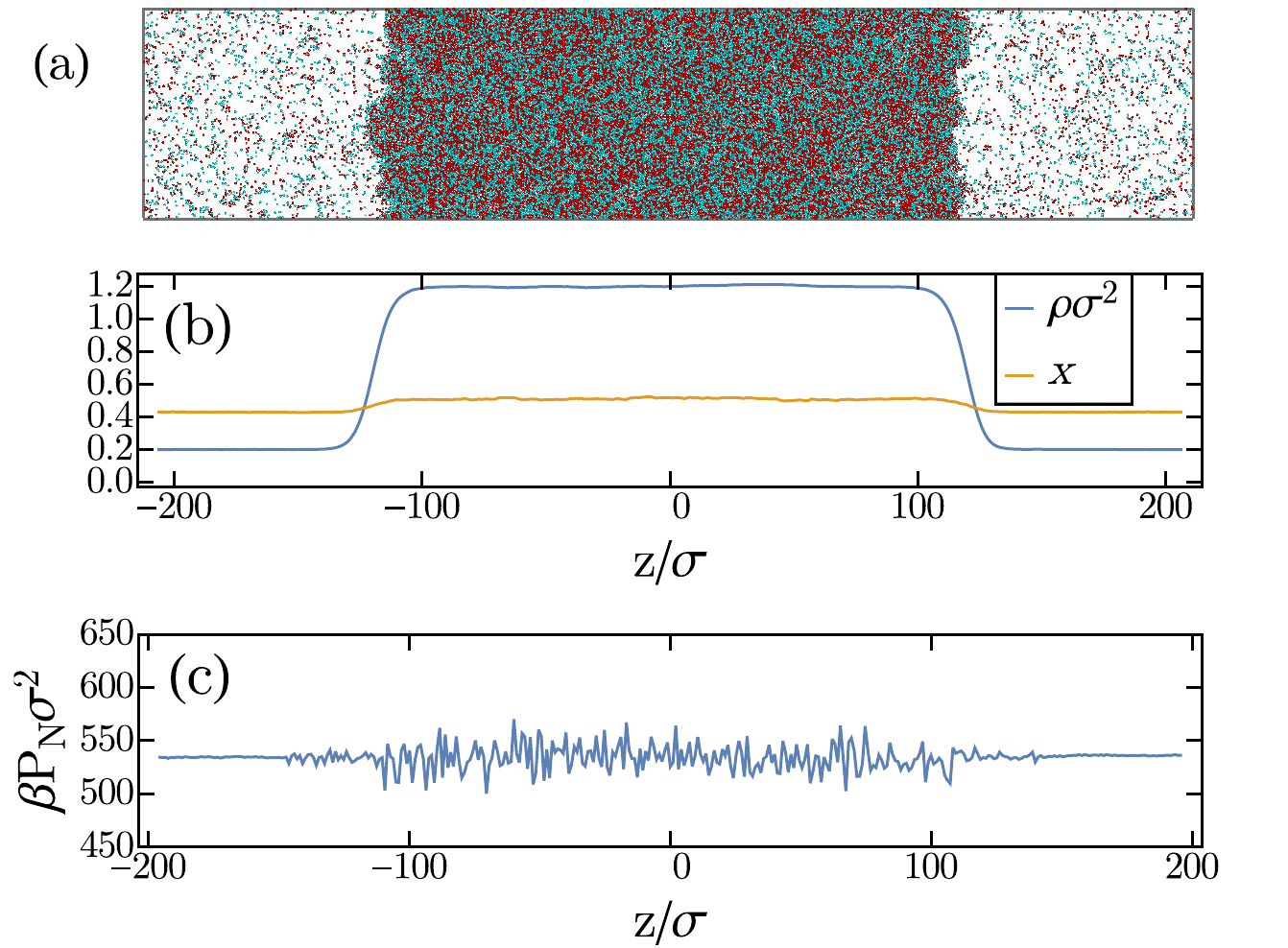} 
\end{center}
\caption{(a) Motility-induced phase separation of an active-active binary mixture with overall composition $x=0.5$ and overall density $\rho\sigma^2=0.75$. Red particles are ``fast" swimmers with self-propulsion force $f_f=160k_BT/\sigma$ and blue particles are ``slow" swimmers with self-propulsion force $f_s=120k_BT/\sigma$. The system contains approximately 30000 particles. (b) The corresponding density ($\rho$) and composition profile ($x$) along the long direction $z$.  (c) The corresponding normal pressure profile along the long direction $z$.}
\label{profiles}
\end{figure*}

So far, we have shown that the gas-liquid coexistence for an active-passive mixture of Lennard-Jones particles can be completely described by the local normal pressure and the reservoir densities. This of course raises the interesting question whether or not such phase coexistence rules can also be found for systems undergoing a motility-induced phase separation, and whether or not they can be used to \textit{predict} the phase diagram. 

To this end, we use the Weeks-Chandler-Andersen potential with the energy scale $\beta\epsilon_{WCA}=6.67$. Here, we consider a two-dimensional active-active mixture with the self-propulsions of fast and slow species being $f_f=160 k_BT/\sigma$ and  $f_s=120k_BT/\sigma$, respectively. We chose such a set of forces for two reasons. First of all, such a choice ensures that even the slow species undergoes motility-induced phase separation  into a gas and a crystal phase (see e.g. the state diagram in Ref.~\cite{redner2013structure}). As a result, we can probe the full spectrum of compositions of the binary mixture. Secondly and more importantly, such high self-propulsions result in relatively fast dynamics for the system, so that our reservoir simulations can access a large number of configurations within reasonable computational time.

We start off by performing direct coexistence simulations between the gas and the crystal phases,  where we simulated approximately 30000 particles in a two-dimensional elongated box with dimensions $L_z=5L_y$. This choice for the dimensions of the box was done such that  two planar interfaces  are created that span the box perpendicular to its long axis \cite{bialke2015negative}. Note that we also observe the formation of gas bubbles in the crystal phase, which have been  reported in Ref.~\cite{bialke2015negative}. 
We performed direct coexistence simulations for system compositions $x\in[0,1]$ with an interval of $0.1$, and total densities $\rho\sigma^2=0.45, 0.6$ and $0.75$. For the highest density we consider, the dimensions of the box are approximately $90 \sigma \times 450 \sigma$. Thus, the short axis of the box is larger than the persistence lengths of the active particles, which are $\beta D_0 f_f/D_r \approx 53.3  \sigma$ and $\beta D_0 f_s/D_r = 40  \sigma$, for the fast and slow species respectively.  The simulations were initiated from a configuration where all particles are part of a hexagonal crystal and ran for approximately  $1500\tau$. We collected data  only for the last $500\tau$. The long running times are necessary for the relaxation of the compositions of the coexisting phases. 

\begin{figure*}
 
\includegraphics[width=0.7\textwidth]{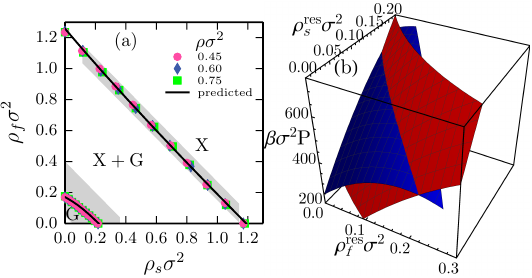}
 
\caption{(a) Direct coexistence results for the phase diagram of the active-active mixture of WCA particles (markers), and the predicted phase diagram (line) in the density-density representation for fast and slow swimmers, as denoted by $\rho_f$ and $\rho_s$, respectively. To predict the binodals we calculated the pressure $P$ and reservoir densities for the fast and slow species, denoted by $\rho_f^{\mathrm{res}}$ and $\rho_s^{\mathrm{res}}$, for a wide range of binary crystal and gas phases (shaded area). Note that X denotes the crystal phase while G denotes the gas phase. (b) Surface plots of the gas (red) and crystal (blue) reservoir densities vs the pressure. From the intersection between the two surfaces we obtain the predicted binodals in (a).}
\label{PhaseDiagWCA}
\end{figure*}

We then measured the local densities of the two species and the normal component of the pressure by dividing  the box into slabs of length $\sigma$ across the long $z$-axis and measuring the corresponding quantities for each slab. Typical results for such measurements as well as a snapshot of the system in direct coexistence are shown in Figure~\ref{profiles}. The results of these direct coexistence simulations are summarized in the phase diagram, as shown in Figure~\ref{PhaseDiagWCA}(a). Clearly, also for this active-active mixture, we observe a collapse of the direct coexistence data onto a single binodal.

As a next step, we predict the phase boundaries by calculating the pressure $P$ and reservoir densities $\rho_f^{\mathrm{res}}$ and $\rho_s^{\mathrm{res}}$ for a wide range of binary crystal and gas phases (shaded area in Figure \ref{PhaseDiagWCA}(a)). 
This involves putting the  binary bulk phase in contact with a reservoir of slow and fast particles separately, for both the crystal and the gas phase. 

To construct the surface that corresponds to the gas in Figure \ref{PhaseDiagWCA}(b), we ran simulations of a gas phase in contact with a reservoir in order to acquire the reservoir densities for fast and slow swimmers. The gas phases  we considered had compositions in the regime $x\in[0.1,0.9]$ and densities $\rho\sigma^2\in[0.05,0.5]$ (see shaded area in Figure \ref{PhaseDiagWCA}(a)). The interval between the points we considered was $0.1$ for the composition and $0.02-0.05$ for the density. We find that for total density $\rho\sigma^2>0.4$ the system spontaneously phase separates so we cannot probe this high-density regime. For each gas phase, which corresponds to a point in the $(x,\rho\sigma^2)$ grid, we ran two simulations, one in contact with a reservoir of fast swimmers and one in contact with a reservoir of slow swimmers. Each simulation provided us with a value for the density of the corresponding reservoir and also a value for the local normal pressure in the binary gas. The values for the pressures were reassuringly in close agreement, since we simulated the same binary phase in both simulations. Thus a pair of such simulations provided us with a point on the hypersurface $(\rho_f^{\rm res},\rho_s^{\rm res},\beta P\sigma^2)$, which is a function of the variables $x$ and $\rho\sigma^2$. As the value of the pressure  $\beta P\sigma^2$ we take the average from the two simulations. This surface can be well fitted by a second degree polynomial. The red surface shown  in Figure \ref{PhaseDiagWCA}(b)  is the fitted polynomial.

For the crystal phase we followed a similar approach. We simulate the binary crystal in contact with a reservoir of slow and fast particles separately. The points we simulated were in the interval of compositions $x\in[0.1,0.9]$ and densities $\rho\sigma^2\in[1,1.30]$, with intervals $0.1$ for the composition and $0.02$ for the density (see shaded area in Figure \ref{PhaseDiagWCA}(a)).  Again, the results of the simulations give us a hypersurface which we again fit with a second degree polynomial.  This polynomial is the blue surface shown  in Figure \ref{PhaseDiagWCA}(b).


\BvdM{Note that in these reservoir simulations, we also applied a short-ranged wall torque to the particles of the confined species that reorients these particles away from the wall. Specifically, particles of the confined species that approach the wall are instantaneously rotated by 180$^\circ$, such that they are oriented towards the bulk of the simulation box. This wall torque was applied in order to minimize the accumulation of particles on the wall. Without this wall torque we observe that for the majority of our reservoir simulations the semi-permeable membrane gets clogged up by a strong adsorption layer of confined particles. As a result exchange between the reservoir and the bulk binary phase is extremely slow such that we need to run the simulation for exceedingly long times to sample the reservoir densities. We have confirmed that the addition of the wall torque does not affect the reservoir density nor the pressure in the bulk phase. 
}

%
%
%
%

Using the requirement of mechanical and chemical equilibrium we accurately construct the phase boundaries from these properties. The intersection of the two surfaces in Figure \ref{PhaseDiagWCA}(b) is the line where the reservoir densities of both species and pressures of the  binary crystal and gas phases are equal. Thus, it should correspond to the binodals of the phase diagram. The lines on each of the two $(\rho_f^{\mathrm{res}},\rho_s^{\mathrm{res}},P)$ surfaces can be converted back into lines in the $x-\rho$ and subsequently the $\rho_f-\rho_s$ representation of the phase diagram. The resulting phase boundaries are drawn in Figure \ref{PhaseDiagWCA}(a)  in order to compare them with the results from the direct coexistence simulations. These ``predicted'' binodals are in very good agreement with the direct coexistence data. We thus have predicted \textit{quantitatively} the phase diagram for an active-active mixture undergoing motility-induced phase separation. This result highlights that also for these systems, which are extremely far from equilibrium, simple coexistence rules are satisfied. Therefore, this result sheds new light on the thermodynamics of systems of active spherical particles.

\section{Conclusions}

In conclusion, we have demonstrated, for the first time, that the phase coexistence of active spherical \BvdM{(torque-free)} particles is fully governed by  mechanical and chemical equilibrium. We have highlighted the generality of our results by applying our technique to two very different active systems. Our results clearly show that phase coexistence for mixtures of active particles can be completely described by the local normal pressure and the reservoir densities per species. Using this requirement of three sets of equal thermodynamic quantities we have quantitatively {\it predicted} phase coexistences for these torque-free active systems. 
\BvdM{We would like to point out that we have restricted ourselves to torque-free active systems as it has been shown that in the presence of torques the pressure is no longer a state function~\cite{solon2015pressure2}. Consequently, for systems with torques one cannot naively follow the route presented in this paper and construct the phase diagram by requiring equal pressure and reservoir densities.
} 

We would like to emphasize that we have introduced a purely numerical method which allows one to 
construct phase diagrams from chemical and mechanical equilibrium without making approximations and without requiring a priori knowledge of the interface between the coexisting phases -- the only interfaces present in our work divide the reservoir from the bulk phase, and are not the same as the ones that appear in the phase diagram coexistences. This numerical observation is intriguing, in particular when compared to theoretical treatments, such as Refs.  \cite{solon2018generalized,hermann2019phase, rodenburg2018ratchet}, where the characteristics of the interface play varying roles.   
%
%
We hope that this work will inspire new theoretical investigations in this direction.

\section{Acknowledgements}   
We acknowledge funding from the Dutch Sector Plan Physics and Chemistry,  and L.F. acknowledges financial   support from the Netherlands Organization for Scientific Research (NWO-VENI Grant No. 680.47.432). We would like to thank Frank Smallenburg for many useful discussions and carefully reading the manuscript. 


\bibliography{ActivePassive}

\end{document}